\documentclass[twocolumn, traditabstract]{aa} % for the abstract without structuration 
\usepackage{graphicx}
\usepackage{txfonts}
\begin{document}
%\setpagewiselinenumbers
%\modulolinenumbers[5]
%\linenumbers
%
   \title{The D/H ratio in the atmospheres of Uranus and Neptune from Herschel PACS observations \thanks{Herschel is an ESA space observatory with science instruments provided by European-led Principal Investigator consortia and with important participation from NASA.}
}

%   \subtitle{I. Overviewing the $\kappa$-mechanism}

\author{H.~Feuchtgruber\inst{1}
\and E.~Lellouch\inst{2}
\and G.~Orton\inst{3}
\and T.~de Graauw\inst{4}
\and B.~Vandenbussche\inst{5}
\and B.~Swinyard\inst{6,7}
\and R.~Moreno\inst{2}
\and C.~Jarchow\inst{8}
\and F.~Billebaud\inst{9,10}
\and T.~Cavali\'e\inst{9,10}
\and S.~Sidher\inst{6}
\and P.~Hartogh\inst{8}
}

\institute{Max-Planck-Institut f\"ur extraterrestrische Physik, Giessenbachstra\ss e, 85748 Garching, Germany \\             
\email{fgb@mpe.mpg.de}
\and LESIA, Observatoire de Paris, 5 place Jules Janssen, F-92195 Meudon, France      
\and Jet Propulsion Laboratory, California Institute of Technology, United States  
\and Alma Observatory, Santiago, Chile         
\and Instituut voor Sterrenkunde, Katholieke Universiteit Leuven, Belgium               
\and Rutherford Appleton Laboratory, Oxfordshire, United Kingdom                        
\and University College London, Gower Street, London, UK
\and Max-Planck-Institut for Solar System Research, Katlenburg-Lindau, Germany  
\and Univ. Bordeaux, LAB, UMR 5804, F-33270, Floirac, France
\and CNRS, LAB, UMR 5804, F-33270, Floirac, France      
%\and Observatoire de Bordeaux, France                                                 
}           

   \date{Draft \today ; received ???; accepted ???}

 \abstract{
Herschel-PACS measurements of the rotational R(0) and R(1) HD lines in the atmospheres of Uranus and Neptune are analyzed
in order to derive a D/H ratio with improved precision for both planets.
The derivation of the D/H ratio includes also previous measurements of the R(2) line by the Short Wavelength Spectrometer on board the 
Infrared Space Observatory (ISO). The available spectroscopic line information of the three rotational transitions is discussed 
and applied in the radiative transfer calculations. 
The best simultaneous fit of all three lines requires only a minor departure from the Spitzer temperature profile of Uranus and a departure limited to 2K from the Voyager temperature profile of Neptune (both around the tropopause).
The resulting and remarkably
similar D/H ratios for Uranus and Neptune are found to be (4.4$\pm$0.4)$\times$10$^{-5}$ and (4.1$\pm$0.4)$\times$10$^{-5}$ respectively. 
Although the deuterium enrichment in both atmospheres compared to the protosolar value
is confirmed, it is found to be lower compared to previous analysis. Using the interior models of Podolak et al. (1995), 
Helled et al. (2011) and Nettelmann et al. (2013), and assuming that complete mixing of the atmosphere and interior occured
during the planets history,  we derive a D/H in protoplanetary ices between (5.75--7.0)$\times$10$^{-5}$ for Uranus and between 
(5.1--7.7)$\times$10$^{-5}$ for Neptune. Conversely, adopting a cometary D/H for the protoplanetary ices between (15-30)$\times$10$^{-5}$, 
we constrain the interior models of both planets to have an ice mass fraction of 14-32\%, i.e. that the two planets are rock-dominated.
 }


\authorrunning{Feuchtgruber et al.}

\titlerunning{D/H in Uranus and Neptune}

\maketitle

\section{Introduction}
Among the light nuclides synthesized during the early evolution of the universe, deuterium is unique in its sensitivity to determine the cosmological density of baryons. 
As early as in the pre-main sequence of stars, deuterium was burned up to $^{3}$He. The gas, which returns into the interstellar matter by stellar outflows and supernova 
explosions, is then free of deuterium. On the other hand, no process is known to produce deuterium, therefore its abundance is decreasing with time, particularly with
progressing star formation. The deuterium abundance as measured today thus provides a lower limit for its corresponding primordial value. 
The determination of accurate D/H ratios in the atmospheres of the Giant Planets has therefore been a longstanding target of research, because their values enable constraining the D/H ratio in the part of our Galaxy where the Solar system formed.
Moreover, the D/H ratio is known to increase in icy grains with decreasing temperature due to ion-molecule and grain-surface interactions (Watson \cite{watson74}; Brown \& Millar \cite{brown89}). Measuring the D/H ratio as a function of heliocentric distance in the Solar system therefore enables probing the temperature of formation of icy grains in the protoplanetary disk (Owen et al. \cite{owen99};  Hersant et al. \cite{hersant01}; Gautier \& Hersant \cite{gautier05}), since the D/H value measured in the atmospheric gas can be linked to the D/H value in the protoplanetary ices as will be shown in section~\ref{discussion}.     

The D/H ratio in the hydrogen of the atmospheres of Jupiter and Saturn is believed to be very close to the protosolar value, because the mass of their cores is negligible with respect to their total mass and because H$_2$ is by far the main constituent of their atmospheres. Consequently, deuterium enrichment of the hydrogen reservoir through ices played only a minor role during their formation. 
However, the atmospheres of Uranus and Neptune are expected to have atmospheres enriched in deuterium. According to Guillot (\cite{guillot99}),  their cores (which in the models are composed of 25\% rock and 60-70\% of ice) make up for more than half of the total mass and mixing of deuterium-enriched icy grains and planetesimals with the hydrogen envelope during 
their formation (Hubbard \& McFarlane \cite{hubbard80}) must have led to a substantially larger D/H ratio in their atmospheres as compared to the protosolar value.    

Molecular hydrogen represents the major fraction of the atmospheres of the Giant Planets ($\sim$85\%). Therefore it is particularly well suited to determine the deuterium abundance from the HD/H$_{2}$ ratio. A number of deuterium abundance determinations from 
infrared observations of CH$_{3}$D have been reported for Uranus and Neptune (de Bergh et al. \cite{debergh86}, de Bergh et al. \cite{debergh90}, Orton et al. \cite{orton92}, Fletcher et al. \cite{fletcher10}, Irwin et al. \cite{irwin12}), however the knowledge of the required isotopic enrichment factor f=(D/H)$_{\mathrm{CH_4}}$/(D/H)$_{\mathrm{H_2}}$, is uncertain (L\'ecluse et al. \cite{lecluse96}).   
Direct observations of rotational far-infrared HD transitions by ground based observatories are difficult because of the opacity of the earth's atmosphere in the relevant wavelength regions. Optical detections of HD have been reported by Trauger et al. (\cite{trauger73}) and Smith et al. (\cite{smith89a}) on Jupiter and Saturn respectively. Similar measurements of HD at visible wavelengths, reported for Uranus (Trafton \& Ramsay \cite{trafton80}) suffer from low signal-to-noise ratio and blending with weak CH$_4$ lines, so they can provide only upper limits in the case of both, Uranus and Neptune (Smith et al. \cite{smith89b}). On the other hand, space based observatories like ISO (Kessler et al. \cite{kessler96}) and Herschel (Pilbratt et al. \cite{pilbratt10}) have access to the mid- to far-infrared rotational (R-branch) lines of HD (Ulivi et al. \cite{ulivi91}, see also Table~\ref{table:1}). While the Voyager/IRIS instrument could not detect these lines due to insufficient spectral resolution, observations by the medium-to-high resolution spectrometers ISO-SWS (de Graauw et al. \cite{degraauw96}) and ISO-LWS (Clegg et al. \cite{clegg96})
resulted in D/H values with significantly improved precision for all four Giant Planets  (Griffin et al. \cite{griffin96}, Encrenaz et al. \cite{encrenaz96}, Feuchtgruber et al. \cite{feuchtgruber99}, Lellouch et al. \cite{lellouch01}). The baseline spectral response
calibration of the ISO-LWS instrument was obtained from Uranus observations. Consequently, this fact prevented a quantitative analysis of the HD R(0) 112~$\mu$m and HD R(1) 57~$\mu$m lines for both Uranus and Neptune. An analysis 
of ISO-LWS measurements of Mars and Callisto gave only inconsistent results (Davies et al. \cite{davies00}), i.e. a line detection at low signal-to-noise with a relative spectral response derived from Callisto and a non-detection at high signal-to-noise with a relative spectral response derived from Mars.

The Herschel-PACS instrument is therefore the first astronomical facility allowing to access the 
information provided by these two lines with high signal-to-noise ratio.    

One major goal of the Herschel Key Program HssO ("Water and Related Chemistry in the Solar system", Hartogh et al. \cite{hartogh09}) was indeed to observe HD lines in the four Giant Planet atmospheres and improve the uncertainties of their respective D/H ratios. First results from HD 
observations on Neptune by the far-infrared spectrometer PACS (Poglitsch et al. \cite{poglitsch10}) on board Herschel have been presented in Lellouch et al. (\cite{lellouch10}). In this work we present a 
combined analysis of the three rotational far-infrared lines of HD detected with ISO-SWS (HD R(2)) and Herschel-PACS (HD R(0) \& HD R(1)) towards a more accurate D/H ratio in the atmospheres of Uranus and Neptune.       

\section{Observations and Data Reduction}
\label{obsandreduction}

PACS spectrometer observations of the HD R(0) and R(1) lines on Uranus and Neptune have been performed in the high spectral sampling density chop-nod mode of the instrument (Poglitsch et al. \cite{poglitsch10}). The details of the observations are summarized in Table~\ref{table2}. The spectrometer spectral resolution $\lambda$/$\delta\lambda$ at the 
two HD line wavelengths of 56.23~$\mu$m and 112.07~$\mu$m is about 2500 and 950, respectively. Each of the lines has been measured both in range scan mode (covering large wavelength ranges) and in line scan mode (short wavelength interval centered on the line). A number of further shallow grating scans covering the full PACS wavelength range have been carried out on both planets as part of the PACS instrument calibration program on their continua, however those measurements did not reach sufficient signal-to-noise ratios on the two HD lines and are not included here.  As part of the Herschel Open Time (OT) program "Variability in Ice Giant Stratospheres: Implications for Radiative, Chemical and Dynamical Processes" led by G. Orton (OT1\_gorton01\_1 program), a number of observations of the HD R(0) and HD R(1) lines have been taken at different longitudes. The HD R(1) line was part of the observing program on Uranus, but not on Neptune. Fortunately this line is seen in the blue spectrometer channel simultaneously to the red channel R(0) line observations on Neptune, however only in grating order 2, at accordingly lower spectral resolution of 1400. These data are included here as well, allowing us to improve significantly on the statistics and to assess observational reproducibility. The data have been extracted from the Herschel science archive and processed up to Level 1 within HIPE 8.0 (see Poglitsch et al. \cite{poglitsch10}). Rebinning and the combination of the two Nod positions has been done outside HIPE by standard IDL tools. The absolute calibration uncertainty of PACS spectrometer data is about 30\% (Poglitsch et al. \cite{poglitsch10}). For an accurate determination of the D/H ratio, all spectra have been divided by their local continua. Thus, absolute calibration errors cancel out and the uncertainties on the line contrast are only driven by the signal-to-noise ratio of the observations. Although the wavelength scale of the observations has been corrected by the Herschel-Target velocity Doppler shift, residual line center shifts remain. The size of the spatial PACS spectrometer pixels is about 9.4"$\times$9.4" and the width of the Herschel telescope PSF (Point Spread Function) ranges from about 6" to 15" within the PACS wavelength range. Consequently Uranus and Neptune can be considered point sources in the context of these observations. The nominal spacecraft pointing uncertainty can move point sources at significantly different positions within the PACS spectrometer slit. As a result, the wavelengths of the spectra may appear slightly shifted, since the nominal calibration applies to the slit center or for extended sources only (Poglitsch et al. \cite{poglitsch10}). Therefore, to prepare for a best-fit analysis, the spectra of all lines have been recentered onto their rest wavelengths.      

Longwards of the R(1) line, around 56.325~$\mu$m, a strong stratospheric H$_2$O emission line is detected. It is however not included in 
this model, as being inconsequential for the D/H determination.           

The ISO-SWS observations of the HD R(2) line at 37.7~$\mu$m have been carried out in 1996-1997 on the two planets at a spectral resolution of 1700. Together with the HD line, the quadrupolar rotational lines of H$_2$, S(0) at 28.22~$\mu$m and S(1) at 17.03~$\mu$m, have been measured to obtain independent constraints for the thermal profile modeling at similar atmospheric pressure levels. Observational details, data reduction and modeling of the HD R(2) line on the two planets are described in Feuchtgruber et al. (\cite{feuchtgruber99}). To facilitate a common modeling scheme with respect to the PACS spectra, these data have also been divided by the continuum.

\begin{table}

\caption{Summary of PACS observations}             % title of Table
\label{table2}      % is used to refer this table in the text
\centering                          % used for centering table
\begin{tabular}{lcccc}        % centered columns (4 columns)
\hline                 % inserts double horizontal lines
%\noalign{\smallskip}
Target & Exposure & HD Line & Date & $\lambda$ range \\    % table heading
 & [sec] & & & [$\mu$m] \\ 
\hline
   \noalign{\smallskip}
   Uranus	& 6941\tablefootmark{a} 							& R(0)		& 24-Nov-09 	& 102-145\\
   Uranus	& 7996\tablefootmark{a} 							& R(1)		& 25-Nov-09 	& 52-62\\
   Uranus	& 2394\tablefootmark{b} 							& R(0), R(1)	& 6-Jul-10 	& 56, 112\\
   Uranus	& 2$\times$1210\tablefootmark{b} 					& R(0), R(1)	& 12-Jan-11 	& 56, 112\\
   Uranus	& 2$\times$1210\tablefootmark{b} 					& R(0), R(1)	& 13-Jan-11 	& 56, 112\\
   Uranus	& 8\tablefootmark{c}$\times$1210\tablefootmark{b} 	& R(0), R(1) 	& 5-Jun-11 	& 56, 112\\
   Neptune	& 6941\tablefootmark{a}							& R(0) 		& 30-Oct-09	& 102-145\\
   Neptune	& 7996\tablefootmark{a}							& R(1) 		& 30-Oct-09	& 52-62\\
   Neptune	& 3168\tablefootmark{b}							& R(0), R(1) 	& 25-May-10 & 56, 112\\
   Neptune	& 8\tablefootmark{c}$\times$730\tablefootmark{b}	& R(0) 		& 5-Jun-11 	& 112\\
   Neptune	& 8\tablefootmark{c}$\times$730\tablefootmark{b}	& R(1) 		& 5-Jun-11 	& 56\\
%\noalign{\smallskip}
\hline            
\end{tabular} \\
\tablefoot{
All observations have been carried out in standard chopped-nodded mode. The R(0) line has been measured in
the 1$^{st}$ grating order and the R(1) line in 3$^{rd}$ one, except the 8 observations of the R(1) line on Neptune, where the line was only seen in the 2$^{nd}$ grating order.     
\tablefoottext{a}{Range scan mode}
\tablefoottext{b}{Line scan mode}
\tablefoottext{c}{Observations taken within the OT1\_gorton01\_1 observing program have been executed several times on the same day at different longitudes to assess longitudinal variability of temperature.}
}

\end{table}

%                                     Two column figure (place early!)
%______________________________________________ 
\begin{figure*}
   \centering
   \includegraphics{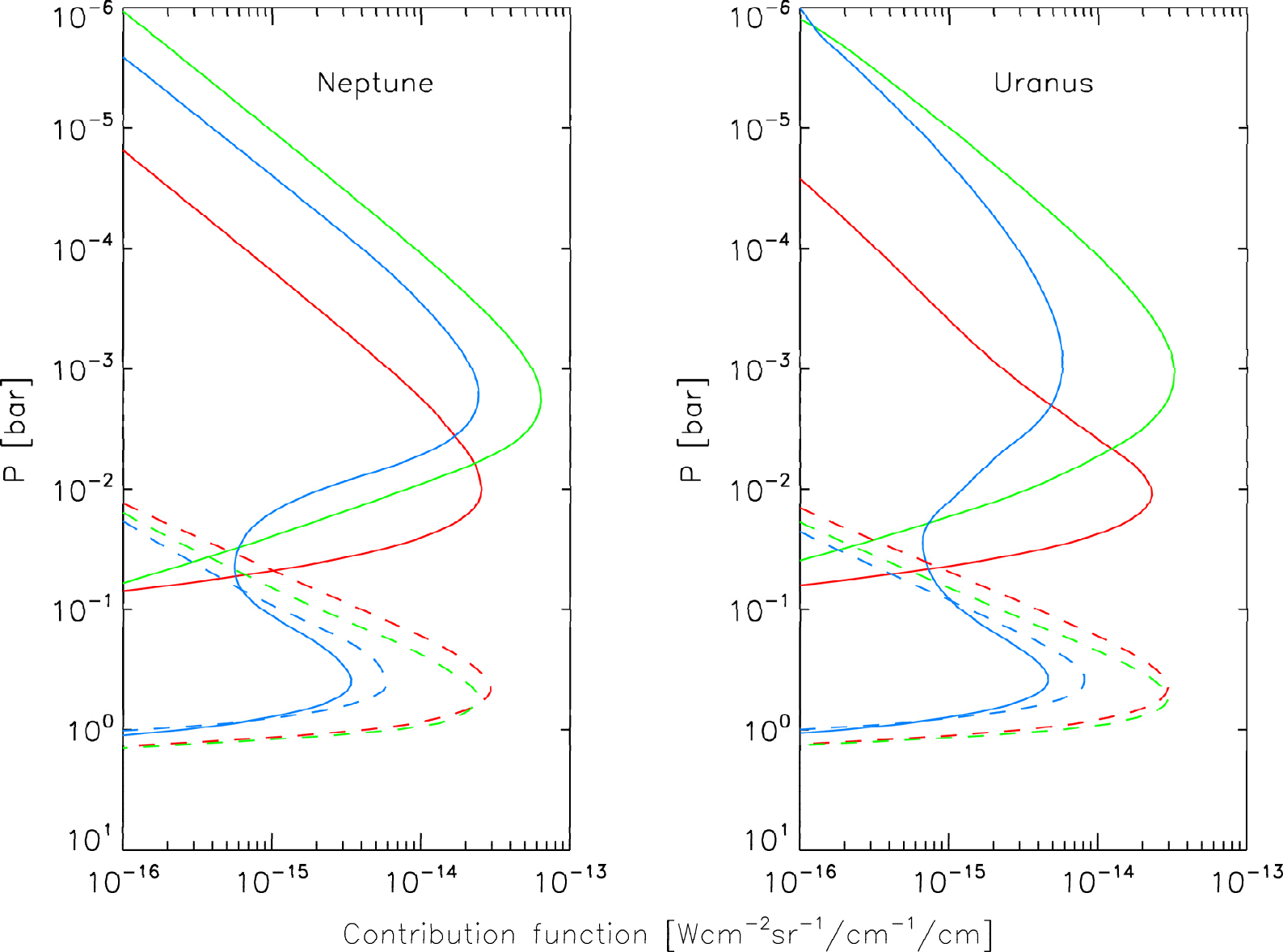}
   \caption{Monochromatic contribution function for the three HD lines (red: R(0); green: R(1); blue: R(2)). Solid: line core;
                   Dashed: continuum}
   \label{fig3}
\end{figure*}

\section{HD Line parameters}
\label{HPlineparameters}

The available information on spectroscopic HD line parameters from the literature has been revisited and a few significant updates with respect to the line parameters used by Feuchtgruber et al. (\cite{feuchtgruber99}) and Lellouch et al. (\cite{lellouch10}) have been worked out. Accurate measurements of the wavelengths of the pure rotational transitions of HD are taken from Evenson et al. (\cite{evenson88}) and 
Ulivi et al. (\cite{ulivi91}) and remain unchanged. The dipole moment $\mu$ of HD is taken now from Table II of Lu et al. (\cite{lu93}) as a mean value across the 4
measured rotational transitions R(0) to R(3). This value of $\mu$=8.21 Debye is then translated into the spectral line intensities given in Table \ref{table:1}.  A 1$\sigma$ uncertainty of $\sim$3\% on the resulting spectral line intensity values is estimated from the scatter in reported dipole moments. Line broadening coefficients $\gamma$ have been measured by Lu et al. (\cite{lu93}) at three different temperatures compared to 295~K only by Drakopoulos \& Tabisz (\cite{drakopoulos87a}) and Drakopoulos \& Tabisz (\cite{drakopoulos87b}). The coefficient $n$ describing the temperature dependence of the half width at half maximum $\gamma$ has been fit to the measured values by: 

 $$
\gamma(\mathrm{T}) = \gamma(\mathrm{T_{ref}})\times(\mathrm{T_{ref}/T)^n}~\mathrm{with}~T_{\mathrm{ref}}=296~\mathrm{K} 
$$

Within the temperature range of $\sim$50-120~K, containing the dominating contributions for HD line modeling in the atmospheres of Uranus and Neptune, the
errors of this fit are $\leq$0.5\% with respect to a linear interpolation of the measurements. For the line parameters $\gamma$, n and the wavenumber shift coefficient $\delta$ of the four transitions, the contributions from the two most significant collisions HD-H$_2$ and HD-He are weighted according to the relative abundance of H$_2$ and He in the 
atmospheres of the outer planets (0.85/0.15).  Updated rotational constants for the HD molecule have been taken from Ulivi et al. (\cite{ulivi91}) and have been included in the code of Ramanlal \&Tennyson (\cite{ramanlal04}) to calculate the temperature dependence of the partition function. The entire set of HD line parameters that has been used in the modeling is given in Table \ref{table:1}. The respective values for the R(3) transition which is not part of this analysis are provided for completeness.   

\begin{table*}
\caption{Adopted HD line parameters for Uranus and Neptune}             % title of Table
\label{table:1}      % is used to refer this table in the text
\centering                          % used for centering table
\begin{tabular}{c c c c c c c}        % centered columns (4 columns)
\hline                 % inserts double horizontal lines
   \noalign{\smallskip}
Line & $\nu$ & Line intensity\tablefootmark{c} & $\gamma$\tablefootmark{c} & $E_{lower}$ & n\tablefootmark{c} & $\delta$\tablefootmark{c} \\    % table heading 
      & $cm^{-1}$   & $cm^{-1}/(molec\ cm^{-2}$) & $cm^{-1}/atm$ & $cm^{-1}$  &  & $cm^{-1}/atm$\\  
\hline
   \noalign{\smallskip}
   R(0) & 89.227950\tablefootmark{a}  & 1.769$\times10^{24}$ & 0.0130 & 0 & -0.232 & 0.0013\\      % inserting body of the table
   R(1) & 177.841792\tablefootmark{b} & 7.517$\times10^{24}$ & 0.0101 & 89.228 & 0.198 & 0.0016\\
   R(2) & 265.241160\tablefootmark{b} & 8.870$\times10^{24}$ & 0.0084 & 267.070 & 0.130 & -0.0045\\
   R(3) & 350.852950\tablefootmark{b} & 4.867$\times10^{24}$ & 0.0086 & 532.311 & -0.030 & -0.0030\\
   \noalign{\smallskip}
\hline            
\end{tabular}\\
\tablefoot{n: Exponent of the temperature dependence of the line half width at half maximum $\gamma$. $\delta$: The  
shift in wavenumber of the line as a function of pressure p [atm]. Both parameters are weighted averages according to the relative
contributions of  HD-H$_2$ and HD-He collisions ($\sim$0.85/0.15).
%$\nu(p)=\nu$+$\delta$$\times p$ \\
\tablefoottext{a}{From Evenson et al. (\cite{evenson88})}
\tablefoottext{b}{From Ulivi et al. (\cite{ulivi91})}
\tablefoottext{c}{Derived from Lu et al. (\cite{lu93})} with T$_{ref}$=296K
}
\end{table*}

%                                     Two column figure (place early!)
%______________________________________________

\begin{figure*}
   \centering
   \includegraphics{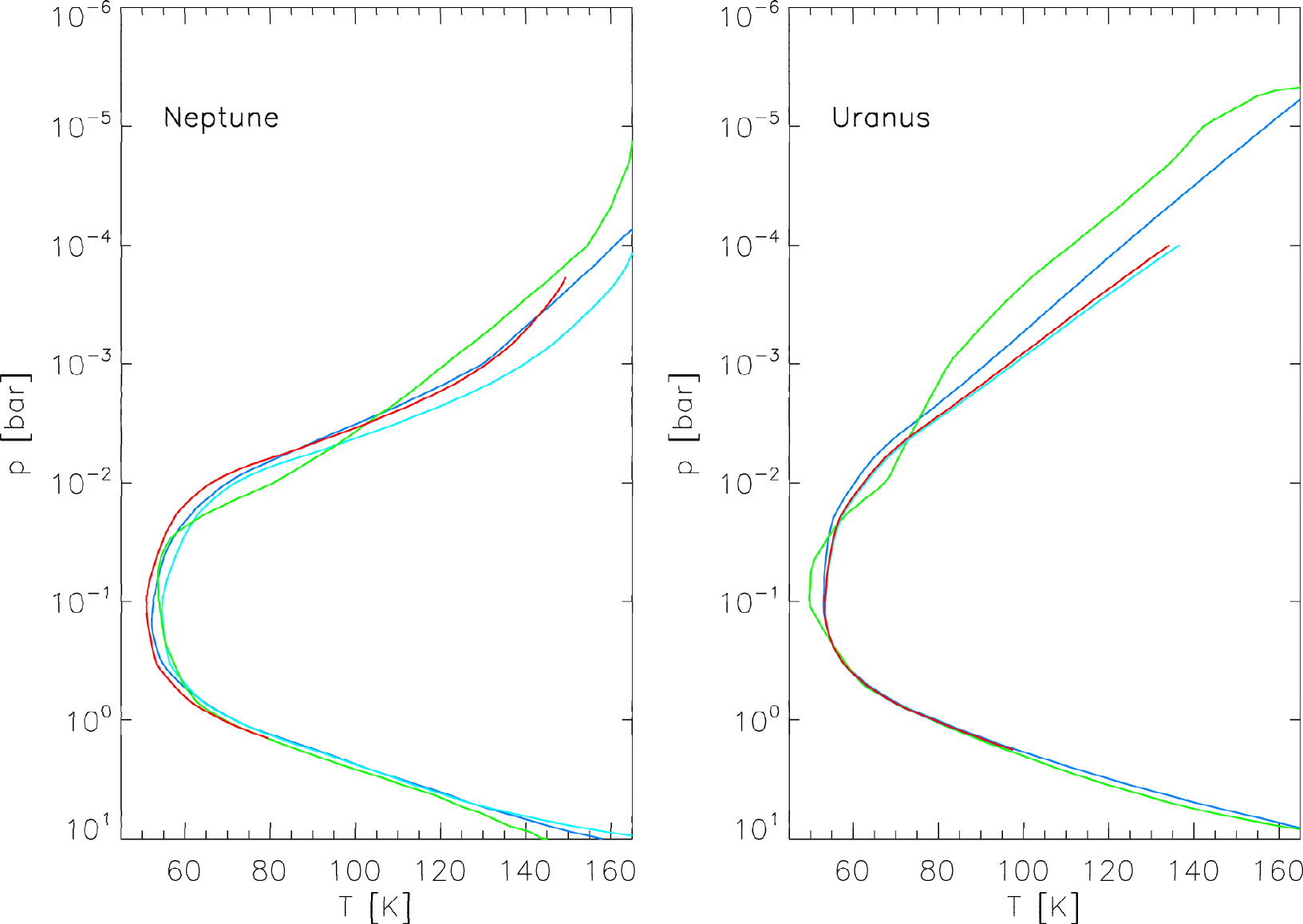}
   \caption{Thermal (p,T) profiles for both planets. Neptune: dark blue = baseline profile (B\'ezard et al. \cite{bezard91}); green = Feuchtgruber et al. (\cite{feuchtgruber99}); light blue = Lellouch et al. (\cite{lellouch10}); red = best fit profile to the three HD lines; 
      Uranus:  blue = baseline profile (Orton et al. \cite{orton13}); green = Feuchtgruber et al. (\cite{feuchtgruber99}); red = best fit profile to the three HD lines; light blue= best fit to R(0) and R(1) lines only}
   \label{fig2}
\end{figure*}

%__________________________________________________________________

\section{Analysis of the D/H ratio}
\label{analysis}

The observations were analyzed by means of a multilayer radiative-transfer model
in which the HD/H$_2$ mixing ratio is assumed to be uniform with altitude. The model includes atmospheric opacities of the three measured HD lines (see Table \ref{table:1})
and collision-induced absorption of H$_2$-H$_2$ (Borysow et al. \cite{borysow85}), H$_2$-He (Borysow et al. \cite{borysow88}), H$_2$-CH$_4$ (Borysow \& Frommhold \cite{borysow86}) and CH$_4$-CH$_4$ (Borysow \& Frommhold \cite{borysow87}). Orton et al. (\cite{orton07}) updated the H$_2$-H$_2$ coefficients of Borysow et al. (\cite{borysow85}), but 
the modifications to far-infrared  absorption were insignificant.

The synthetic spectra were calculated monochromatically, 
integrated over all viewing angles of the planets, and then convolved with the instrumental profile (FWHM = 120, 215, 315 km/s) at the respective wavelengths and grating orders. Monochromatic contribution functions for the line 
centers and their adjacent continua (Figure~\ref{fig3}) indicate the range of layers within the thermal profile dominating the continuum divided spectra of the two planets. The line-to-continuum ratios of the ensemble of the three HD lines are sensitive to atmospheric pressure levels from $\sim$10$^{-4}$ bar to $\sim$1.5 bar. 
An initial thermal profile (p,T) of Uranus has been taken from Orton et al. (\cite{orton13}), representing a best fit to CH$_4$ and CH$_3$D emission spectra from Spitzer IRS data.
The initial profile for Neptune is from B\'ezard et al. (\cite{bezard91}), whose tropospheric part relevant for this work originates from the Voyager radio occultation experiment (Lindal et al. \cite{lindal90}). These baseline profiles for the two planets do not allow to obtain a good match of all three observed HD lines with the model. Together with the HD/H$_2$ mixing ratio, these thermal profiles are adjusted by a $4^{th}$ order polynomial (in log(p)) minimizing the least squares to fit all the continuum divided spectra of the three HD transitions
simultaneously by the model. Figure~\ref{fig2} shows the best fit thermal profile for the two planets, together with the initial input profiles and earlier work
for comparison. Combining the Herschel data 
of the HD lines with Spitzer data for a joint modeling is deferred to the future.

In the case of Uranus, the resulting profile is in remarkable agreement with Feuchtgruber et al. (\cite{feuchtgruber99}) and Orton et al. (\cite{orton13}) for pressures  $\ge$200~mbar. 
The profile of Feuchtgruber et al. (\cite{feuchtgruber99}) is about 5~K colder at 70~mbar, 5~K warmer at 10~mbar and about 15~K colder for pressures lower than 1~mbar than our best fit profile. With 
respect to Orton et al. (\cite{orton13}), the new profile is slightly warmer (up to 2~K) between 100~mbar and 5~mbar, for lower pressures
the new profile becomes continuously warmer up to 10~K at 0.1~mbar however. For Neptune, the baseline profile from B\'ezard et al.  (\cite{bezard91}) and Lindal et al. (\cite{lindal90}) required a slight cooling of 1-2~K starting at 5~mbar towards higher pressure levels. At lower pressures, the differences are then negligible. The profile of Lellouch et al. (\cite{lellouch10})
is warmer at all levels, from about 2~K at 1~bar to 4~K at the tropopause up to around 10~K at 1~mbar.  

In the case of Neptune, the sub-Earth latitude changed only from about -27.3$^\circ$ to -28.3$^\circ$ between 1996 and 2010, justifying the approach of combining the ISO-SWS data with Herschel-PACS data for a joint modeling by
one thermal profile. However the change of sub-Earth latitude for Uranus went from -46.6$^\circ$ in 1996 to +13.5$^\circ$ in 2010. VLA mapping observations of Uranus at centimeter wavelengths are reported by Hofstadter \& Butler (\cite{hofstadter03}) and Hofstadter et al. (\cite{hofstadter11}),
which indicate spatial changes to the thermal profile at pressures $\ge$1 bar within the period of interest. In order to verify whether a possible change in the disk-averaged thermal profile between the epochs of the ISO-SWS and PACS observations may affect our analysis on Uranus, we derived also a thermal profile by a best fit to the R(0) and R(1) lines only. The respective D/H for fitting only these two lines is slightly larger (4.7$\times$10$^{-5}$) but still within the 1$\sigma$ uncertainty, however for such a thermal profile the model over-predicts the R(2) line already by 30\%. The resulting profile (see Fig.~\ref{fig2}) is essentially the same for pressures larger than 100mbar compared to the profile derived from all three lines. Small departures ($\le$2~K) occur only at lower pressures, confirming that our analysis is not affected by seasonal changes when including all three lines in the calculations. 

\begin{figure}

   \centering
     \includegraphics[width=8.4cm]{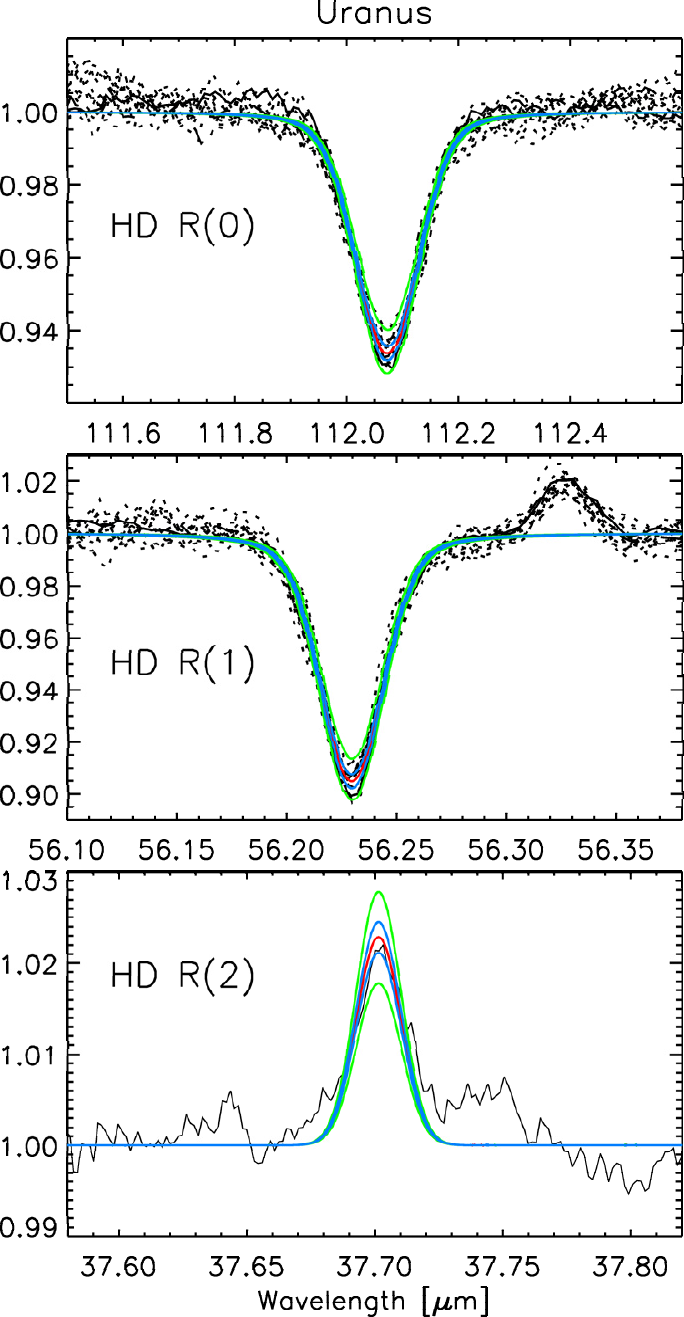}
      \caption{Observed and synthetic Uranus spectra. Black: continuum divided spectra (solid: range scan; dashed: HssO line scan; dotted: OT line scans) and best fit model (red) with D/H=$4.41\times10^{-5}$ . 
               Synthetic spectra for different D/H ratios but for the same thermal profile to illustrate the sensitivity to this model parameter are shown as green solid lines for D/H=$3.4\times10^{-5}$ and D/H=$5.4\times10^{-5}$ (3$\sigma$) and blue solid lines for D/H=$4.07\times10^{-5}$ and D/H=$4.75\times10^{-5}$ (1$\sigma$). The spectral line around 56.33~$\mu$m is caused by stratospheric H$_2$O emission and is not included in the model. The R(2) line has been measured by ISO-SWS.}
 \label{fig4}
\end{figure}

\begin{figure}
 
   \centering
     \includegraphics[width=8.4cm]{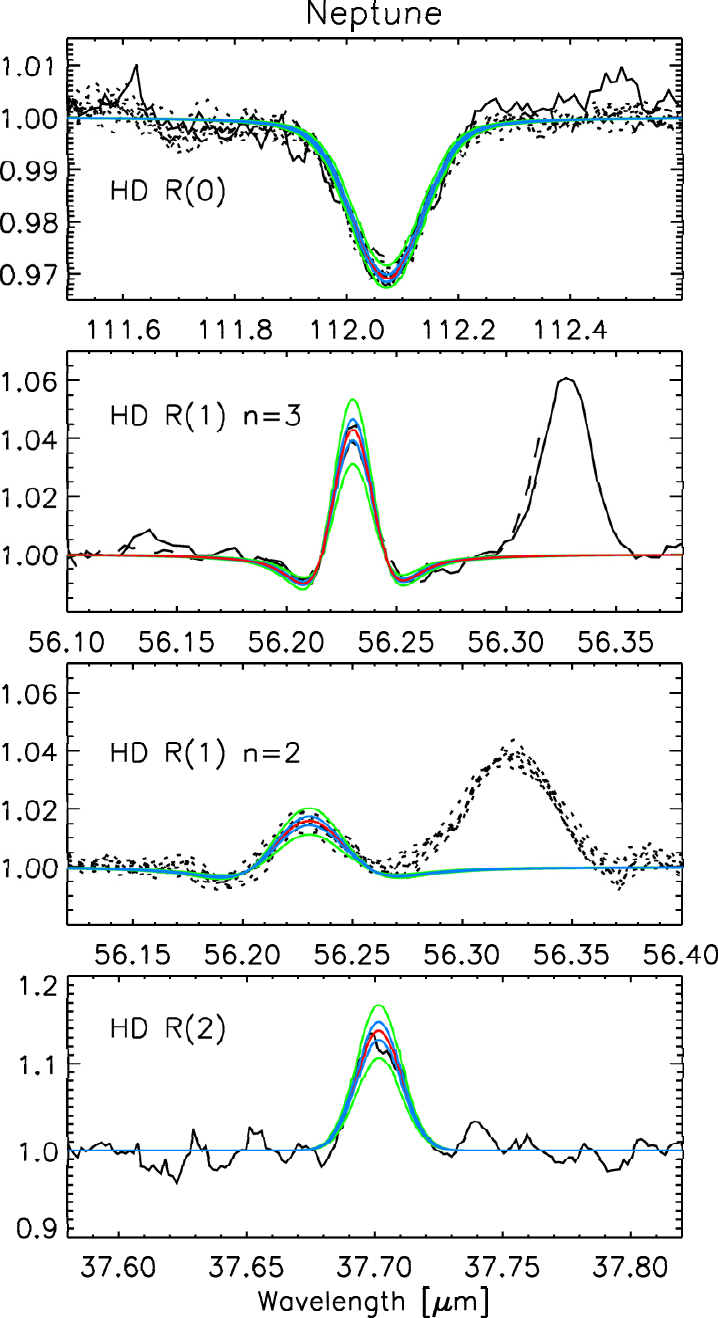}
      \caption{Observed and synthetic Neptune spectra. Black: continuum divided spectra (solid: range scan; dashed: GT line scan; dotted: OT line scans) and best fit model (red) with D/H=$4.08\times10^{-5}$ . 
              Synthetic spectra for different D/H ratios but for the same thermal profile to illustrate the sensitivity to this model parameter are shown as green solid lines for D/H=$3.09\times10^{-5}$ and D/H=$5.07\times10^{-5}$ (3$\sigma$) and blue solid lines for D/H=$3.75\times10^{-5}$ and D/H=$4.41\times10^{-5}$ (1$\sigma$). The R(1) line has been observed in two different grating orders (n) and accordingly at different instrumental resolution. The strong spectral line around 56.33$\mu$m is caused by stratospheric H$_2$O emission and is not included in the model. The R(2) line has been measured by ISO-SWS.}
   \label{fig1}     
\end{figure}

Figures~\ref{fig4} and~\ref{fig1} show all the observed spectra and the calculated models for Uranus and Neptune respectively. The peak-to-peak scatter in line-to-continuum ratios between different PACS observations is about 19\% (14 observations) and 17\% (10 observations) for the R(0) transitions on Uranus and Neptune respectively. However, this 
scatter is mainly due to variations in spacecraft pointing offsets, which may cause slight instrumental profile variations and therefore variations of the peak contrast.  A variation of the best-fit D/H ratios by $\sim\pm1\times10^{-5}$ using the same thermal profiles is matching already well to the extremes within the R(0) and R(1) observations. However, at the same time, variations by this amount appear incompatible with the measured spectra of the R(2) lines. The D/H ratio which matches the extremes within all observations of the R(0) and R(1) lines can be considered as the 3$\sigma$ statistical error. Translated to 1$\sigma$ uncertainties we get D/H=$(4.41\pm0.34)\times10^{-5}$ for Uranus and D/H=$(4.08\pm0.33)\times10^{-5}$ for Neptune.

However, on top of the statistical error from the number of independent observations, there are also systematic uncertainties on the spectroscopic line parameters of HD. Line intensities (3\%, $1\sigma$), broadening parameter and its temperature coefficient, and the wavenumber-shift parameter have independent and transition specific uncertainties of a few percent, which may affect the calculations of the D/H ratio either way. We therefore add a $5\%$ uncertainty in quadrature to our statistical error and finally quote D/H=$(4.4\pm0.4)\times10^{-5}$ and D/H=$(4.1\pm0.4)\times10^{-5}$ for Uranus and Neptune respectively. 
Note finally that a little warming or cooling of the best fit thermal profiles by $\pm$1K leads to synthetic spectra which depart already by $\sim1\sigma$ for all observed lines on Uranus and slightly more than $1\sigma$ for the R(0), R(2) and about $2\sigma$ for the R(1) measurements on Neptune. Therefore, error bars due to thermal profiles uncertainties can be neglected compared to those due to the scatter in observed line contrasts.
 
These D/H values can be compared to results on the deuterium content in the methane reservoirs of the two planets.
Using the isotopic enrichment factors from L\'ecluse et al. (\cite{lecluse96}) of $f=1.68\pm0.23$ and $f=1.61\pm0.21$ for Uranus and Neptune respectively with:
$${
f = 
\mathrm{{(D/H)_{CH_4} \over (D/H)_{H_2}} }
}$$  
and 
$${
\mathrm{CH_3D/CH_4 = 4\times (D/H)_{CH_4}}
}$$

we get CH$_3$D/CH$_4$ (Uranus) = $(2.96 {{+0.71}\atop{-0.64}})\times10^{-4}$ and CH$_3$D/CH$_4$ (Neptune) = $(2.64{{+0.64}\atop{-0.56}})\times10^{-4}$, in good agreement  
with recent results by Irwin et al. (\cite{irwin12}) (CH$_3$D/CH$_4$ = $(2.9 {{+0.9}\atop{-0.5}})\times10^{-4}$ for Uranus) and Fletcher et al. (\cite{fletcher10}) (CH$_3$D/CH$_4$ = $(3.0 {{+1.0}\atop{-1.0}})\times10^{-4}$ for Neptune).

\section{Discussion}
\label{discussion}

Our new measurement of the D/H ratio in H$_2$ in Uranus and Neptune can be combined
with a model of their internal structure to constrain the D/H in their protoplanetary ices.
Following the approach proposed by L\'ecluse et al. (\cite{lecluse96}) and also adopted by Feuchtgruber et al. (\cite{feuchtgruber99}), 
(D/H)$_{\mathrm{ices}}$ can be expressed as:

$${
\mathrm{(D/H)_{ices} = {(D/H)_{planet} - x_{H_2} (D/H)_{proto} \over (1 - x_{H_2})}}
}$$

where (D/H)$_{\mathrm{planet}}$ is the bulk D/H ratio in the planet, taken equal to its value in
the fluid envelope (D/H)$_{\mathrm{envelope}}$. This assumes that the atmosphere and interior of the planet 
have been fully mixed, i.e. that high-temperature equilibration of deuterium between hydrogen 
and ices has occurred during the planet's history. This assumption of global mixing 
is central to our analysis. Formation models (Pollack \& Bodenheimer \cite{pollack89}) suggest that the  planetary envelopes mixed in early
stages, but whether this was true also for core material is admittedly unknown. Current Giant Planets may not be fully convective, 
especially Uranus (see Podolak et al. \cite{podolak95}, Guillot \cite{guillot05}), and it is not known whether this state is primordial
or not. Another assumption of the model is
that (D/H)$_{\mathrm{envelope}}$ is equal to the 
D/H value we determine in H$_2$. This assumes that the atmospheric deuterium content is largely 
dominated by H$_2$, with negligible contribution from heavy H-bearing species (e.g. H$_2$O).
This hypothesis is further discussed below.

For the protosolar D/H value, we adopt (D/H)$_{\mathrm{proto}}$ = (2.25$\pm$0.35)$\times$10$^{-5}$, based
on the analysis of ISO-SWS measurements on Jupiter (Lellouch et al. \cite{lellouch01}). For the volumetric
ratio x$_{\mathrm{H_2}}$ = n$_{\mathrm{H_2}}$/(n$_{\mathrm{H_2}}$+n$_{\mathrm{H_2O}}$)  of H$_2$, we initially used the interior models of Podolak et al. (\cite{podolak95}). For Neptune, Podolak et al. (\cite{podolak95}) considered two model variants, once with a ``canonical" density in the ice shell, and one
with a density reduced by 20~\%.
These models provide values (expressed in Earth masses) for the gas (M$_{\mathrm{H_2+He}}$), ice (M$_{\mathrm{ice}}$),
and rock (M$_{\mathrm{rock}}$) components of the planet. Note that all the Podolak et al. (\cite{podolak95}) models assumed a solar rock-to-ice ratio ($\sim$2.5), i.e. an ice mass ratio in the heavy element component (F = M$_{\mathrm{ice}}$/(M$_{\mathrm{ice}}$ + M$_{\mathrm{rock}}$)) equal to  0.715. Assuming solar composition, i.e. H$_2$/(H$_2$+He) = 0.747 by mass (consistent with Helled et al. \cite{helled11}), x$_{\mathrm{H_2}}$ can be expressed as:

$${ 
x_{\mathrm{H_2}} = {1 \over 1 + {(1-f_{\mathrm{H_2}}) \over (m_{\mathrm{H_2O}}/m_{\mathrm{H_2}})\times f_{\mathrm{H_2}})}}
}$$ 

where  
$${
f_{\mathrm{H_2}} = {0.747 M_{H_2+He} \over 0.747 M_{H_2+He} + M_{ice}}
}$$

is the mass ratio of H$_2$ and m$_{\mathrm{H_2O}}$ and m$_{\mathrm{H_2}}$ are the molar masses of H$_2$O and H$_2$ respectively (i.e. 18g and 2g). The x$_{\mathrm{H_2}}$ values and the inferred (D/H)$_{\mathrm{ices}}$ are given in 
Table~\ref{protoplanetaryices} (note that the x$_{\mathrm{H_2}}$ values are slightly different from those given in 
Feuchtgruber et al. (\cite{feuchtgruber99}), due to a minor mistake in that paper).

We also used the more recent Uranus and Neptune models of Helled et al. (\cite{helled11}). These models
propose empirical pressure-density models tuned to match the planetary radii, masses, gravitational
coefficients J$_2$ and J$_4$ and solid rotation periods determined by Voyager. The models are 
then interpreted in terms of bulk composition, i.e. the mass fractions of hydrogen (X), helium (Y),
and heavy elements (Z). Two limiting cases are considered for heavy elements, being either pure rock
(represented for definiteness by SiO$_2$) or pure ice (represented by H$_2$O). Furthermore, two variants (I and II) are considered regarding the radial 
distribution of the heavy elements within the planetary interior. Obviously the pure SiO$_2$ 
cases are excluded from the point of view of the D/H ratio, since they would lead to a 
protosolar D/H. We thus considered here the H$_2$O cases. By definition, these models have F = 1. In this case, f$_{\mathrm{H_2}}$ is simply given
as f$_{\mathrm{H_2}}$ = X / (X+Z). 

Even more recently, interior models of Uranus and Neptune were updated 
by Nettelmann et al. (\cite{nettelmann13}). These authors provide results based on full sets of three-layer interior models  combined with different solid-body rotation periods, gravitational data and physical equations of state (2 models
for Uranus and 3 models for Neptune). Values for x$_{\mathrm{H_2}}$, f$_{\mathrm{H_2}}$, and 
resulting (D/H)$_{\mathrm{ices}}$ from all their models are given in Table~\ref{protoplanetaryices}.

\begin{table} %[ht]
   \caption{Inferred D/H in protoplanetary ices}
  \label{protoplanetaryices}
   \begin{tabular}{lccc}
   \hline
    & & &\\[-1.0ex]
%Table 3. Inferred D/H in protoplanetary ices
 
Model                    &               f$_{\mathrm{H_2}}$ &     x$_{\mathrm{H_2}}$       &    (D/H)$_{\mathrm{ices}}$  \\
\hline
  & & &\\[-1.0ex]
%\cite{podolak95} & & &\\
Uranus$^a$                                     & 0.108 & 0.521       & (6.75+1.1/-1.2)$\times$10$^{-5}$ \\
Neptune 1$^a$                                & 0.133 & 0.581       & (6.7$\pm$1.4)$\times$10$^{-5}$ \\    
Neptune 2$^a$                                & 0.055 & 0.343       & (5.1$\pm$0.75)$\times$10$^{-5}$ \\
\hline
  & & &\\[-1.0ex]    
%\cite{helled11}  & & &\\
Uranus I  (H$_2$O)$^b$                   & 0.087 & 0.463       & (6.25$\pm$1.05)$\times$10$^{-5}$ \\
Uranus II (H$_2$O)$^b$                   & 0.065 & 0.387       & (5.75$\pm$0.9)$\times$10$^{-5}$ \\ 
Neptune I (H$_2$O)$^b$                  & 0.082 & 0.444       & (5.58$\pm$1)$\times$10$^{-5}$ \\   
Neptune II (H$_2$O)$^b$                 & 0.074 &  0.417       & (5.42$\pm$0.95)$\times$10$^{-5}$ \\
\hline
& & &\\[-1.0ex]    
%\cite{Nettelmann13} & & &\\
Uranus 1$^c$				  &	0.116	&	0.541	&	(6.94$\pm$1.3)$\times$10$^{-5}$				\\
Uranus 2$^c$					   &	0.097	&	0.492	&	(6.48$\pm$1.1)$\times$10$^{-5}$				\\	
Neptune 1$^c$					   &	0.150	&	0.614	&	(7.04$\pm$1.6)$\times$10$^{-5}$				\\
Neptune 2a$^c$					   &	0.178	&	0.661	&    (7.71$\pm$1.9)$\times$10$^{-5}$				\\
Neptune 2b$^c$					   &	0.124	&	0.560	&	(6.46$\pm$1.4)$\times$10$^{-5}$				\\
\hline
\end{tabular}
\phantom{}\\[1.0ex]
%\smallskip
Models from: (a) Podolak et al. (\cite{podolak95}), (b) Helled et al. (\cite{helled11}) \\
(c) Nettelmann et al. (\cite{nettelmann13})\\ 
Volumetric ratio: x$_{\mathrm{H_2}}$ = n$_{\mathrm{H_2}}$/(n$_{\mathrm{H_2}}$+n$_{\mathrm{H_2O}}$) \\
Mass ratio:  f$_{\mathrm{H_2}}$ = M$_{\mathrm{H_2}}$/(M$_{\mathrm{H_2}}$ + M$_{\mathrm{ice}}$)

\end{table}

For all the interior models considered, Table~\ref{protoplanetaryices} indicates that the D/H 
values for the  protoplanetary ices of Uranus and Neptune are consistently  
4-6 times smaller than those found in the water-ice reservoirs of Oort-cloud comets 
($\sim$(2-3)$\times$10$^{-4}$, see Bockel\'ee-Morvan et al. \cite{bockelee12}), and still a factor of
2-3 lower compared to the D/H in Earth's oceanic water (1.5$\times$10$^{-4}$),
carbonaceous chondrites (1.4$\times$10$^{-4}$) and the Jupiter-family comet 103P/Hartley
2 (1.6$\times$10$^{-4}$, Hartogh et al. \cite{hartogh11}).

It is somewhat unexpected to find that proto-uranian and proto-neptunian ices
are much less D-rich than cometary ices, warranting a brief discussion of the above hypothesis and derivation. 
A first issue is that (D/H)$_{\mathrm{envelope}}$ may not be equal to (D/H)$_{\mathrm{H_2}}$. This
situation occurs if the planet envelope is heavily enriched in D/H-rich volatiles. However,
even for an extreme water enrichment within the atmosphere (e.g. O/H = 440 times solar, as 
advocated for Neptune by Lodders \& Fegley (\cite{lodders94})), the (D/H)$_{\mathrm{envelope}}$ is increased only by a 
factor 1.24 ((L\'ecluse et al. \cite{lecluse96}), leading to only a $\sim$30 \% increase of (D/H)$_{\mathrm{ices}}$, by far
insufficient for reconciliation with cometary values. The 
second possibility is that the assumption of complete mixing is not valid. In a state of incomplete mixing,
the derived values of (D/H)$_{\mathrm{ices}}$ as per Table \ref{protoplanetaryices} would represent lower limits of the D/H in the icy cores.
Note however that the indistinguishable values of (D/H) in H$_2$ (within error bars) in Uranus and 
Neptune (which still differ radically in terms of their internal energy sources and therefore presumably of
their convective state) suggest that the current incomplete 
convection in Uranus' interior may not be important in this respect. Therefore, we feel that this scenario
is also not obviously promising.

The (D/H)$_{\mathrm{ices}}$ values we infer for the protoplanetary ices correspond to a modest enrichment factor
f$\sim$2-3 over the protosolar value. Compared to the most pristine (i.e. D-rich)
solar system material (i.e. the D/H-rich component of the LL3 meteorites, having f=35) 
or even to typical cometary material (f$\sim$7-20), this implies that the protoplanetary 
ices have been considerably reprocessed in the solar nebula. Yet evolutionary models 
(Kavelaars et al. \cite{kavelaars11}) accounting for radial turbulent mixing within the nebula predict
much larger enrichments (f=14-20) at the estimated 12-15 AU formation distance of 
Uranus and Neptune. Along with the recently revealed diversity of the D/H ratio in comets 
(Hartogh et al. \cite{hartogh11}, Bockel\'ee-Morvan et al. \cite{bockelee12}) and the apparent absence of correlation 
of the values with the estimated formation distance of these comets (i.e., the Kuiper-Belt
vs Oort cloud families) our result of a low D/H in the Uranus and Neptune original ices 
would illustrate the limitation of these evolutionary models. Note also that given the error bars, we are unable to find any significant
difference between the D/H in the proto-uranian and proto-neptunian ices (see Table \ref{protoplanetaryices}), so searching for a correlation between the D/H and the formation distance (as predicted by the evolutionary models), is precluded.

The idea that the protoplanetary ices should necessarily have a deuterium
content equal to that measured in cometary water may
however be challenged by the work of Alexander et al. (\cite{alexander12}). They find a
linear correlation between the D/H and C/H measured in
a set of carbonaceous chondrites (CC) from the CM and CR groups that
experienced different degrees of aqueous alteration.
This suggests that the hydrogen isotopic composition in these bodies
results from the mixing between hydrated silicates
and organic matter, and extrapolating the relationship to C/H = 0 should
therefore give the isotopic composition of water.
This approach provides (D/H)$_{\mathrm{H_2O}}$  $\sim$9$\times$10$^{-5}$ for CMs
(but $\sim$17$\times$10$^{-5}$ for CRs).
In a second approach, the authors correct the D/H measured in other types
of chondrites from the contribution of organic
material and infer that in addition to CMs, the water compositions of the
CIs, CO, CV and Taglish Lake meteorites are less
deuterium-rich than comets, with D/H generally below
$\sim$10$\times$10$^{-5}$. As chondrites are fragments of main-belt
asteroids (and in particular, CCs are generally associated with C-type
asteroids, which may have formed in the same region
as comets (Walsh et al. \cite{walsh11}) ), this is thus indication that there may have existed material
with ``low" D/H in water ice originating from the formation
region of comets. It might therefore not be irrelevant to relate the low
values we infer for the proto-uranian/proto-neptunian
ices ((4--9)$\times$10$^{-5}$, see Table \ref{protoplanetaryices}) to such material. However,
a complication in this scenario is to understand why so far no comets have been
observed to exhibit such a low D/H, and what is the origin of the cometary
D/H enhancement compared to this value.

Coming back to the scenario where cometary ices are representative of protoplanetary ices,
all the above calculations rely on interior models of Uranus and Neptune, which
are not well constrained (see e.g. a discussion on the effect of the uncertainty of the
rotation period by Podolak \& Helled \cite{podolak12}). It is therefore worthwhile to ``invert" the problem,
i.e. assume some value of (D/H)$_{\mathrm{ices}}$ and constrain the interior structures. As the pure SiO$_2$ 
models of Helled et al. (\cite{helled11}) cannot explain a (D/H)$_{\mathrm{H_2}}$ larger than the protosolar value, while
the pure H$_2$O models lead to too small values for (D/H)$_{\mathrm{ices}}$, it is clear that intermediate 
models (i.e. a mix of ice and silicates) are needed. We therefore searched for the ice mass 
ratio in the heavy element component (F = M$_{\mathrm{ice}}$/(M$_{\mathrm{ice}}$ + M$_{\mathrm{rock}}$), targeting 
(D/H)$_{\mathrm{ices}}$ = 1.5$\times$10$^{-4}$ or 3$\times$10$^{-4}$. For a given input value of F, the X, Y, Z 
values were interpolated from Table 3 of Helled et al. (\cite{helled11}), considering the average of 
cases I and II. Z$_{\mathrm{ice}}$ and Z$_{\mathrm{rock}}$ are then given by F$\times$Z and (1-F)$\times$Z, 
respectively, and (D/H)$_{\mathrm{ices}}$ was obtained as before, using now f$_{H_2}$ = X/(X+Z$_{\mathrm{ice}}$).
Finally, the knowledge of X, Z$_{\mathrm{ice}}$ and Z$_{\mathrm{rock}}$ permits us to derive the mass of H, 
the total mass of O, and the mass of O contained in the ice, which can be translated
into O/H ratios. Results are given in Table~\ref{interiormodels}, where the O/H ratios are expressed
in mass, volume, and in the enhanced factor over the solar value (assuming volume solar O/H = 
4.9$\times$10$^{-4}$; Asplund et al. (\cite{asplund09})). Of course the precise values given in 
Table~\ref{interiormodels} are somewhat dependent on the simplified description, attached to the
Helled at al. \cite{helled11} models, that all the ice is in the form of H$_2$O and all the rock in the form of SiO$_2$.
And again, results are subject to the validity of the complete mixing hypothesis: for incomplete mixing, the F values reported
in Table \ref{interiormodels} would represent lower limits to the actual ice mass fraction.

\begin{table} %[ht]
   \caption{"Inferred" interior models}
  \label{interiormodels}
   \begin{tabular}{lcccc}
   \hline
    & & & &\\[-1.0ex]
            &                                  Uranus   & &           Neptune &  \\
Target (D/H)$_{\mathrm{ice}}$  ($\times$10$^{-5}$)     		&   15   &    30   &      15  &   30 \\
F = M$_{\mathrm{ice}}$/(M$_{\mathrm{ice}}$+M$_{\mathrm{rock}}$)    	&  0.32  &   0.15  &     0.28 &  0.14 \\
Z = Z$_{\mathrm{ice}}$+Z$_{\mathrm{rock}}$                                    &0.81  & 0.79 & 0.80 & 0.78  \\
\hline
 & & & &\\[-1.0ex]
f$_{\mathrm{H_2}}$                                                                                  &   0.354        &    0.570       &    0.396        &   0.609       \\
x$_{\mathrm{H_2}}$                                                                                    &   0.831    &     0.923      &     0.855       &  0.933        \\
\hline
 & & & &\\[-1.0ex]
O/H ratio (total  O)  						    	&           &           &            &        \\
Mass          									&  3.08  &   2.71  &    2.91  &   2.58 \\
Volume          								&  0.193 &   0.169 &    0.182 &   0.161 \\  
$\times$ solar\tablefootmark{a}     				&  393   &   345   &    372   &   329 \\
\hline
 & & & &\\[-1.0ex]
O/H ratio (O in ice)      	&           &           &            &        \\
Mass         									&  1.35  &   0.619 &   1.160  &   0.532 \\
Volume          								& 0.084  &   0.039 &    0.073 &   0.033     \\
$\times$ solar\tablefootmark{a}     				& 172    &   79      &    148   &   68 \\
\hline
\end{tabular}
%\phantom{}\\[1.0ex]
%\smallskip
\tablefoot{
\tablefoottext{a}{solar O/H volume ratio = 4.9$\times$10$^{-4}$, Asplund et al. (\cite{asplund09})}
}
\end{table}

The models in Table~\ref{interiormodels} have F = 0.14-0.32. In other words, we infer that 68-86 \%
of the heavy component consists of rock and 14-32 \% is made of ice. Therefore, unlike in the Podolak et al. (\cite{podolak95}) models 
and in the ice (H$_2$O) version of the Helled et al. (\cite{helled11}) models, we find that Uranus and 
Neptune interiors might be more rocky than icy. This behaviour is similar to the case of Pluto, which based on
the body density ($\sim$2 g cm$^{-2}$), has an estimated rock mass fraction of about 0.7 (Simonelli \& Reynolds \cite{simonelli89}). All the above
models have Z = 78--81 \% per mass. With Uranus' and Neptune's masses equal to  14.5 and 17.1 Earth
masses, this gives $\sim$11.6 M$_{\oplus}$ of heavy elements for Uranus and $\sim$13.5 M$_\oplus$ for Neptune. This is to be compared with
the findings by Owen \& Encrenaz (\cite{owen03}, \cite{owen06}) who similarly used the enrichment in heavy elements in the Giant Planets (mostly measured in carbon)
to estimate the mass of the ``SCIP" (solar-composition icy planetesimals) within each planet. For Neptune, Owen \& Encrenaz (\cite{owen06})
find a SCIP mass of (13$\pm$3) M$_\oplus$, fully consistent with our value. The agreement is worse at Uranus, where Owen \& Encrenaz (\cite{owen06})
find (8.5$^{+2.5}_{-2.0}$) M$_\oplus$.

The total O/H ratio is 329-393 times solar, but when
only O from H$_2$O is considered, the O/H ratio is 68-172 times solar. Independent constraints on the atmospheric O/H
ratio have been inferred from the measured CO mixing ratio, a disequilibrium species whose tropospheric abundance is sensitive both 
to the vigor of vertical mixing from the deep atmosphere and to the O/H ratio (CO is produced from H$_2$O from the net
reaction CH$_4$ + H$_2$O $^{\longrightarrow}_{\longleftarrow}$ CO + 3H$_2$.)  To explain the $\sim$1~ppm CO abundance initially
measured in Neptune's troposphere (Marten et al. \cite{marten93}, Guilloteau et al. \cite{guilloteau93}) Lodders \& Fegley (\cite{lodders94}) invoked 
a 440 times solar O/H ratio in Neptune's deep atmosphere, using a solar O/H = 7.4$\times$10$^{-4}$. Rescaling this to O/H = 
4.9$\times$10$^{-4}$ gives O/H = 660 times solar, which is 4.5-9 times higher than we infer. We conclude that the Lodders \& Fegley (\cite{lodders94}) models are inconsistent with our D/H measurement, although reconciliation may be possible if the CO abundance
is actually overestimated by a considerable factor. As a matter of fact, subsequent observations  (Lellouch
et al. \cite{lellouch05}, Hesman et al. \cite{hesman07}, Luszcz-Cook \& de Pater \cite{luszcz13}) all indicate that Neptune CO has two distinct components and
that its tropospheric abundance is lower than previously thought. Luszcz-Cook \& de Pater (\cite{luszcz13}) in particular, determined 
a much smaller ($\sim$0.1 ppm) CO tropospheric abundance. However, by reassessing the Lodders and Fegley (\cite{lodders94}) model, especially in terms
of (i) the limiting reaction steps and (ii) the characteristic mixing time, they still find that 
that a global O/H enrichment of at least 400, and likely more than 650, times the protosolar value is required to explain their measured CO
abundance. Therefore, the discrepancy with our estimate of the atmospheric O/H remains.

Finally, we note that the O/H volume ratios for the H$_2$O component are in the range 
0.033-0.084, i.e. H$_2$O / H$_2$ = 0.07 - 0.17 in the atmosphere. This induces
only a minor correction to (D/H)$_{\mathrm{H_2}}$ (L\'ecluse et al. \cite{lecluse96}), i.e. (D/H)$_{\mathrm{envelope}}$ = (1.04-1.09)$\times$(D/H)$_{\mathrm{H_2}}$
and does not impact any of the above conclusions.

\section{Conclusions}

Herschel-PACS spectrometer observations combined with previous ISO-SWS observations of the three lowest energy rotational lines of HD on Uranus and Neptune have been analyzed. Applying latest spectroscopic line parameters for the HD R(0), R(1) and R(2) lines in multilayer atmospheric radiative transfer calculations, a least-squares fit of synthetic spectra to the continuum divided measurements results in D/H values of (4.4$\pm$0.4)$\times$10$^{-5}$ and (4.1$\pm$0.4)$\times$10$^{-5}$ ($1\sigma$) for Uranus and Neptune respectively. 
The simultaneous modeling of the three measured HD lines requires only small modifications to thermal profiles (p,T) known from earlier work on both planets. The new D/H values, although somewhat smaller than in previous analysis, confirm the enrichment of deuterium in the atmospheres of Uranus and Neptune compared to the protosolar nebula. 
Based on published interior models in which the largest fraction of the heavy
elements is in the form of ice (70-100 \%), and assuming that complete mixing of the atmosphere and interior occured
during the planets history, the required D/H in the protoplanetary
ices responsible for this enrichment is significantly lower than known
from any water/ice source in the solar system. A possible solution to this unexpected
result is that the interiors of Uranus and Neptune are actually dominated by 
rock. For icy material in the form of H$_2$O and rocky material in the form of SiO$_2$, an ice mass fraction of only 14-32 \% is found.
Still, a complete interpretation of the now accurate values of Uranus and Neptune D/H would greatly be aided by a more definite 
picture of their internal structures. The latter would benefit from improved gravity, shape and rotation data, that could be obtained from planetary orbiters.

\begin{acknowledgements}
We thank Bruno B\'ezard for important advice on the HD line parameters, and Dominique Bockel\'ee-Morvan for discussions  on D/H
in meteorites and comets.
G. Orton carried out a part of this research at the Jet Propulsion Laboratory, California Institute of Technology, under a constract with NASA.
T. Cavali\'e wishes to thank the Centre National d'\'Etudes Spatiales (CNES) for funding. 
F. Billebaud acknowledges pluri-annual funding from the Programme National de Plan\'etologie (PNP) of CNRS/INSU.
PACS has been developed by a consortium of institutes led by MPE (Germany) and including UVIE (Austria); KUL, CSL, IMEC (Belgium); CEA, OAMP (France); MPIA (Germany); IFSI, OAP/AOT, OAA/CAISMI, LENS, SISSA (Italy); IAC (Spain). This development has been supported by the funding agencies BMVIT (Austria), ESA-PRODEX (Belgium), CEA/CNES (France), DLR (Germany), ASI (Italy), and CICYT/MCYT (Spain).
\end{acknowledgements}

\end{document}